\documentclass{appolb}
\usepackage{epsfig}
\usepackage{axodraw-1}
\def\lsim{\mathrel{\raise.3ex\hbox{$<$\kern-0.8em\lower1ex\hbox{$\sim$}}}}
\def\gsim{\mathrel{\raise.3ex\hbox{$>$\kern-0.7em\lower1ex\hbox{$\sim$}}}}

\def\eg{{\it e.g.}}

\begin{document}

\title{Scalar gluons and Dirac gluinos at the LHC\thanks{Presented by J. Kalinowski at the 9th Hellenic School and Workshops on Elementary Particle Physics and Gravity, August 30 - September 6, 2009, Corfu, Greece, and
 the XXXIII International Conference of Theoretical Physics MATTER TO THE DEEPEST, September 11-16, 2009, Ustro\'n, Poland.}%
}
\author{S.~Y.~Choi$^1$, J.~Kalinowski$^2$, J.~M.~Kim$^3$ and E.~Popenda$^4$
\address{ $^1$ Dept. Physics and RIPC, Chonbuk National University, Jeonju 561-756, Korea \\
          $^2$ Faculty of Physics, U.~Warsaw, PL-00681 Warsaw, Poland, and \\
          CERN, Theoretical Physics, CH-1211 Geneva 23, Switzerland\\
          $^3$ Phys. Inst. and Bethe CPT, U.~Bonn, D-53115 Bonn, Germany \\
          $^4$ Inst. Theor. Physik, KIT, D-76131 Karlsruhe, Germany \\
          } }
               \maketitle
\begin{abstract}
The hybrid N=1/N=2 supersymmetric model predicts scalar gluons (sgluons) as SUSY partners of the Dirac gluino. Their strikingly distinct phenomenology at the CERN Large Hadron Collider is discussed.
\end{abstract}

\prepNo{CERN-PH-TH/2009-214}

\PACS{12.60.Jv, 14.80.Ly}

\section{Introduction}

Among many propositions for the physics beyond the Standard Model (SM),
supersymmetry (SUSY) is generally considered most elegant and respected. It can accommodate or explain some of the outstanding problems of the SM. In fact this is the only mathematically consistent UV completion of the SM that stabilizes the gap between electroweak and Planck scales. It provides the gauge coupling unification, radiative
electroweak symmetry breaking, a candidate for dark matter (DM), offers new ideas on the
matter-antimatter asymmetry etc.

The simplest N=1 supersymmetric extension calls for each SM particle a sparticle that differs in spin by half a unit. The  Lagrangian must be supplemented by SUSY breaking terms that keep unseen superpartners out of the current experimental reach while retaining all goodies of the model: renormalizability and  perturbatively stable hierarchy of scales. With ever improving experimental constraints on SUSY breaking parameters, mainly from flavor and Higgs physics, the model building of successful SUSY breaking scenarios becomes more and more difficult.

However, the successes of SUSY do not rely on its simplest realization. In fact, non-minimal realizations may ameliorate some of the above problems. For example, Dirac gauginos (in contrast to Majorana in N=1) forbid some of the couplings and may lead to additional suppression of loop contributions with gauginos
in flavor-changing processes. Such scenarios can be based on D-term supersymmetry breaking models \cite{fnw,npt} or continuous R-symmetries \cite{rsymm}.

A Dirac gaugino requires additional fermionic degrees of freedom. They can be provided by adding  chiral super-multiplets in the adjoint representations of the corresponding gauge group, as realised in N=2 SUSY   \cite{secondSusy}. When extending to N=2, the additional supersymmetry also requires the introduction of mirror matter superfields. In order to avoid chirality problems the N=2 mirror
(s)fermions have to be assumed to be heavy. Thus effectively we consider a hybrid
model that expands to N=2 only in the gaugino sector~\cite{CDFZ}.

Here we will present elements of the phenomenology of the scalar partner of the Dirac gluino (sgluon), as worked out in a recent paper~\cite{our}; see also~\cite{ours}. For an independent analysis of sgluons at the LHC, we refer to Ref.~\cite{TT}.

\section{Basics of the N=1/N=2 hybrid model}

If gluinos are seen at the LHC one of the main experimental goals will be to verify if they are Majorana or Dirac fermions. To address properly this issue a theoretical model is needed that allows a smooth Dirac/Majorana transition, as for example the hybrid model constructed in Ref.~\cite{CDFZ}.

In N=1 gluinos are Majorana fields with two degrees of freedom to match gluons in the color-octet vector super-multiplet.
To provide the two additional degrees of freedom for Dirac fields, the usual N=1 gluon/gluino $\{g^a,\tilde{g}^a\}$
super-multiplet $W^a_{3\alpha}=\tilde{g}^a_\alpha +D^a\theta_\alpha+ {(\sigma^{\mu\nu})_\alpha}^\beta \theta_\beta G^a_{\mu\nu}+\ldots$ ($a=1,\ldots,8$)
 is supplemented  by an additional N=1 color-octet chiral super-multiplet
  $\Phi^a=\sigma^a+\sqrt{2}\theta^\alpha\tilde{g}'^{a}_\alpha+ \theta\theta F^a$
  of extra gluinos $\tilde{g}'^{a}$ and scalar
$\sigma^a$ fields to form a vector hyper-multiplet of N=2 SUSY. Similarly, the electroweak sector, not to be discussed here,  is supplemented by additional SU(2)$_L$ and U(1)$_Y$ super-multiplets \cite{Belanger:2009wf}.

\subsection{Dirac/Majorana gluinos}
Standard  $\tilde{g}$ and new gluinos $\tilde{g}'$ couple minimally to the gluon field
\begin{equation}
    {\mathcal{L}}_{\rm SQCD} \ni g_s {\rm Tr}\,
    ( \overline{\tilde{g}} \gamma^{\mu} [{g}_{\mu}, {\tilde{g}}] +
          \overline{\tilde{g}'} \gamma^{\mu} [{g}_{\mu}, {\tilde{g}'}] ) \,,
  \label{eq:gluino}
\end{equation}
as required by the gauge symmetry. Here $g_s$ denotes the QCD coupling, the fields $g_\mu,\, \tilde{g}, \, \tilde{g}'$ are color-octet matrices (\eg\ for the gluon ${g}_\mu =
\frac{1}{\sqrt{2}} \lambda^a g^a_\mu$ with the Gell-Mann matrices $\lambda^a$);
$\tilde{g}$  and $\tilde{g}'$  are two 4-component Majorana spinor fields.  Quark and squark fields interact only with the standard gluino,
\begin{equation} \label{eq:qcd-yuk}
   \mathcal{L}_{\rm SQCD} \ni
   -  g_s (\,\overline{q_L}  \tilde{g} \, \tilde{q}_L
                 - \overline{q_R}  \tilde{g} \, \tilde{q}_R
                 + {\rm h.c.}) \,,
\end{equation}
since only their mirror partners (assumed to be heavy) couple to $\tilde{g}'$, as required by N=2 SUSY.

Soft supersymmetry breaking generates masses for the gluino fields $\tilde{g}$
and $\tilde{g}'$,
\begin{equation}
   {\mathcal{M}}_g =\left(\begin{array}{cc}
   M_3 & M^D_3 \\
   M^D_3 & M'_3
\end{array}\right).
\label{eq:gluinomass}
\end{equation}
Diagonal terms are induced by the individual Majorana mass parameters $M_3$ and $M_3'$ while an
off-diagonal term corresponds to the  Dirac mass. It can arise from the D term, $\int d^2\theta \frac{\sqrt{2}}{M_0}W^{'\alpha} W^a_{3\alpha} \Phi^a+h.c.$ When a hidden sector spurion field $W'_\alpha$ gets a vacuum expectation value, $\langle W'_\alpha\rangle =D'\theta_\alpha$, the Dirac mass is generated, $M^D_3=D'/M_0$.

Diagonalization of (\ref{eq:gluinomass}) gives rise to two Majorana mass eigenstates, $\tilde{g}_1$ and $\tilde{g}_2$,
with masses $m_1$ and $m_2$. There are two limiting cases of interest: in the limit $M_3' \to \pm
\infty$ the standard Majorana gluino is recovered;  in
the limit of vanishing Majorana mass parameters $M_3$ and $M_3'$ with
 $M^D_3\neq 0$, the mixing is
maximal and the two Majorana gluino states are degenerate. One can draw a smooth path interpolating these two limiting cases by introducing an auxiliary parameter $y\in [-1,0]$ and parameterizing $M'_3=y\, M^D_3/(1+y)$, $M_3=-y\, M^D_3$. With $M^D_3=m_{\tilde{g}_1}$ kept fixed, the parameter $y$ allows for a continuous transition from $y=-1$, where the standard limit is reached with one Majorana gluino (the second being infinitely heavy), to $y=0$ that corresponds
to two degenerate Majorana fields combined to a Dirac gluino
\begin{equation}
\tilde{g}_D=\tilde{g}_R+\tilde{g}'_L.
\end{equation}

\begin{figure}\begin{center}
\epsfig{figure=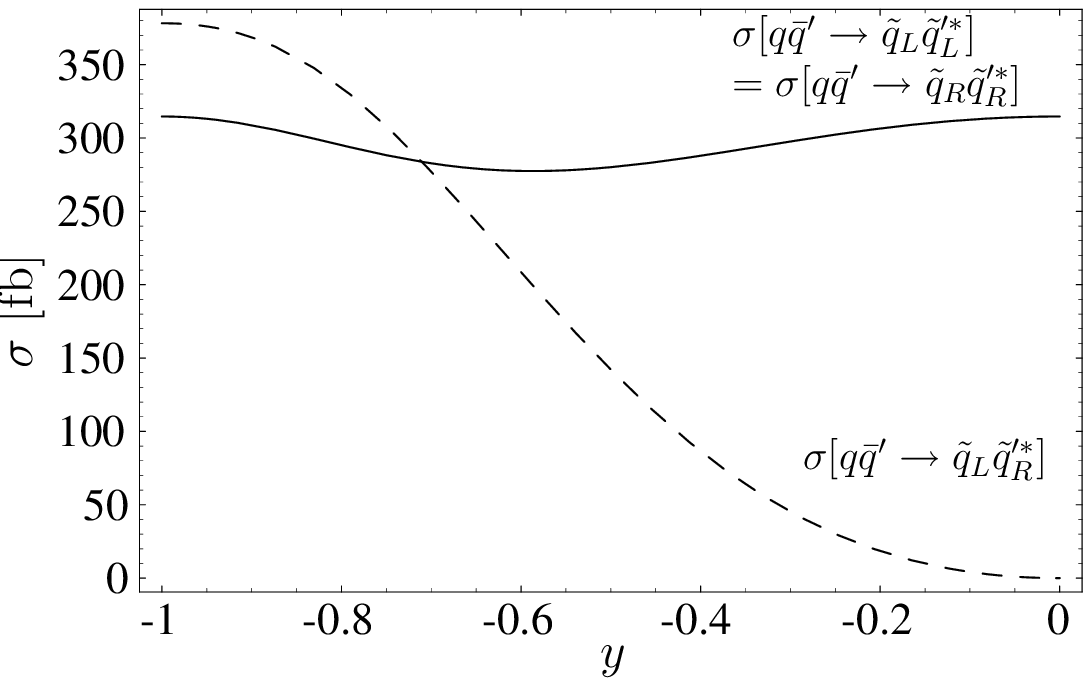, width=6cm}\quad \epsfig{figure=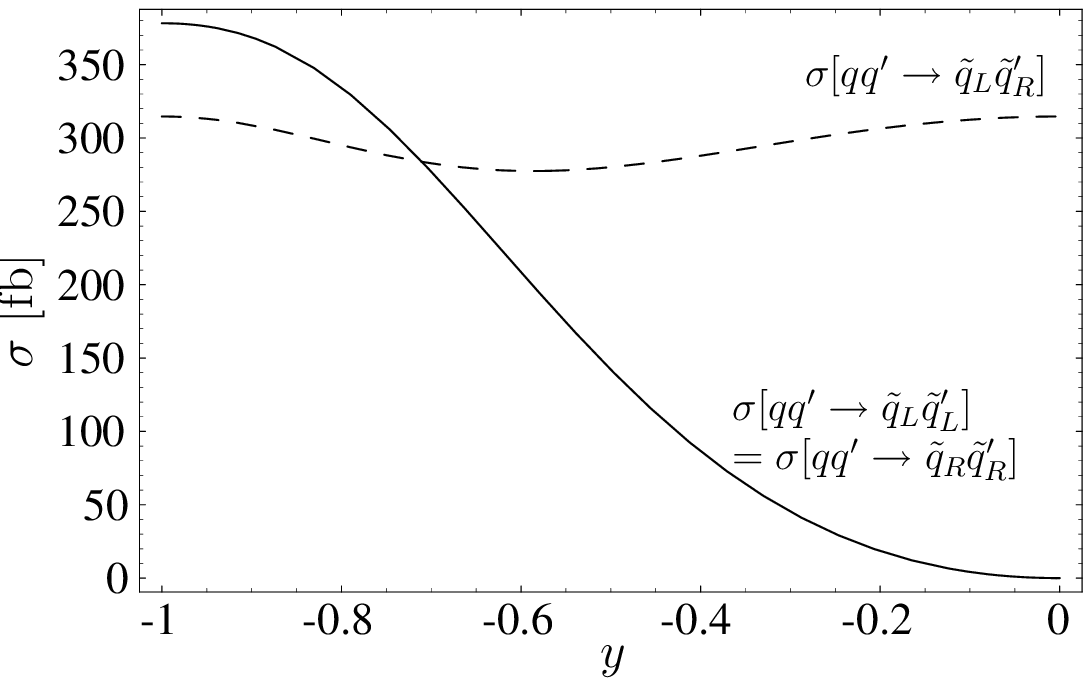, width=6cm}
\vspace{-2ex}
\caption{\it Cross sections for different-flavor squark production in
$q\bar q'\to\tilde{q}\tilde{q}'^\ast$ (left) and $qq'\to\tilde{q}\tilde{q}'$ (right)   as a
function of the Dirac/Majorana control parameter~$y$.}
\label{fig:D/M}\end{center}
\end{figure}

The phenomenology of Dirac gluinos is characteristically different from Majorana. The  detailed discussion can be found in Ref.~\cite{CDFZ}, from which we borrow an example for illustration. Fig.~\ref{fig:D/M} shows the partonic cross sections $q\bar q'\to\tilde{q}\tilde{q}'^\ast$ (left) and $qq'\to\tilde{q}\tilde{q}'$ (right) for different-flavor squark production, in both cases  mediated by the gluino $t$-channel exchange,  plotted as a
function of a Dirac/Majorana control parameter~$y$ (for partonic center-of-mass energy $\sqrt{s}=2000$ GeV,
$m_{\tilde q} = 500$ GeV and $m_{\tilde{g}_1} = 600$ GeV).
\newcommand{\qw}{\ensuremath{\tilde{q}}}
We see that $\sigma[q \bar{q}' \to \qw_L \qw_R'^\ast]$ and $\sigma(qq' \to \tilde q_L \tilde q_L')= \sigma(qq' \to \tilde q_R \tilde q_R') $ are non-zero in the Majorana limit
but vanish in the Dirac, while $\sigma[q \bar{q}' \to \qw_L \qw_L'^\ast]=\sigma[q \bar q' \to \qw_R \qw_R'^\ast]$ and $ \sigma(qq' \to \tilde q_L \tilde q_R')$ are non-zero for any $y$ and
 reach the same values in both limits. For equal-flavor quark-antiquark scattering the additional gluino $s$-channel exchange must be added to the $t$-channel exchange diagrams.

\subsection{Sgluons}

The gluinos $\tilde{g}'^{a}$ are accompanied by a color-octet complex scalar field $\sigma^a$.
In parallel to the case of degenerate Majorana gluinos combined
to the Dirac, the real and imaginary components of the scalar field $\sigma$ will be assumed degenerate with the sgluon mass denoted by $M_\sigma$~\cite{our}.

Although sgluons are R-parity even, they cannot be singly produced at tree level in gluon-gluon or quark-antiquark collision. This is so because they couple only in pairs to gluons, $\sigma\sigma^*g$ and $\sigma\sigma^*gg$, as required by the gauge symmetry, and singly only to Dirac gluino pairs via the Yukawa-type gauge interaction
\begin{eqnarray}
{\cal L}_{\tilde{g}_D\tilde{g}_D\sigma}
   &=& -\sqrt{2} i\, g_s\, f^{abc}\, \overline{\tilde{g}^a_{DL}}\,
        \tilde{g}^b_{DR}\, \sigma^c
       +{\rm h.c.},\label{eq:sglgl}
\end{eqnarray}
where $f^{abc}$ are the SU(3)$_C$ structure constants. It is interesting to note that even the loop-induced $\sigma gg $ coupling due to gluino exchange vanish as a consequence of Bose symmetry since the coupling is even in momentum space but odd,  $\sim f^{abc}$,
in color space.\footnote{Since the sgluon couples only to two different Majorana gluinos, eq.~(\ref{eq:sglgl}), while gluons always couple to the same pair, eq.~(\ref{eq:gluino}), the coupling of the octet sgluon to any number of gluons via the gluino loop is forbidden.}

When SUSY is broken spontaneously, the Dirac gluino mass generates, via the super-QCD $D$ term, a scalar coupling between $\sigma$ and squark pair~\cite{fnw}
\begin{equation} \label{eq:sqq}
{\cal L}_{\sigma\tilde{q}\tilde{q}} =  -
 g_s M_3^D\,  \sigma^a \frac{\lambda^a_{ij}} {\sqrt{2}}
  \sum_q \left( \tilde q_{Li}^* \tilde q_{Lj} - \tilde q_{Ri}^* \tilde q_{Rj}
  \right) + {\rm h.c.}\,.
\end{equation}
Note that the  $L$ and $R$ squarks contribute with
opposite signs. Since squarks couple to gluons and quarks, the loop diagrams with squark/gluino
\begin{figure}[t]
\begin{center}
\begin{picture}(300,60)(0,70)
\Text(0,95)[r]{$(a)$}
\Text(15,95)[c] {$\sigma$}
\Text(100,95)[c] {$\tilde{q}$}
\DashLine(25,95)(50,95){3}
\DashLine(50,95)(90,115){3}
\DashLine(90,75)(50,95){3}
\DashLine(90,115)(90,75){3}
\Text(125,115)[l]{ $g$}
\Text(125,75)[l]{ $g$}
\Gluon(90,115)(120,115){2}{5}
\Gluon(90,75)(120,75){2}{5}
\Text(150,95)[r]{ $\sigma$}
\Text(200,100)[c]{ $\tilde{q}$}
\DashLine(155,95)(180,95){3}
\DashArrowArcn(200,95)(20,180,0){3}
\DashArrowArcn(200,95)(20,0,180){3}
\Gluon(220,95)(250,115){2}{4}
\Gluon(220,95)(250,75){2}{4}
\Text(255,115)[l]{ $g$}
\Text(255,75)[l]{ $g$}
\end{picture}
\begin{picture}(300,60)(0,70)
\Text(0,95)[r]{$(b)$}
\Text(15,95)[r]{ $\sigma$}
\Text(70,115)[c]{ $\tilde{g}_D$}
\Text(95,95)[l]{ $\tilde{q}$}
\DashLine(25,95)(50,95){3}
\Line(50,95)(90,115)
\Line(90,75)(50,95)
\DashLine(90,75)(90,115){3}
\Text(125,115)[l]{ $q$}
\Text(125,75)[l]{ $\bar{q}$}
\ArrowLine(90,115)(120,115)
\ArrowLine(120,75)(90,75)
\Text(150,95)[r]{ $\sigma$}
\Text(200,115)[c]{ $\tilde{q}$}
\Text(225,98)[l]{ $\tilde{g}_D$}
\DashLine(155,95)(180,95){3}
\DashLine(180,95)(220,115){3}
\DashLine(220,75)(180,95){3}
\Line(220,75)(220,115)
\ArrowLine(220,115)(250,115)
\ArrowLine(250,75)(220,75)
\Text(255,115)[l]{ $q$}
\Text(255,75)[l]{ $\bar{q}$}
\end{picture}
\end{center}
\caption{\it Generic diagrams for  the effective $\sigma gg$ (a) and $\sigma q\bar{q}$ (b) vertices  with
             squark/gluino loops.}
\label{fig:loops}
\end{figure}
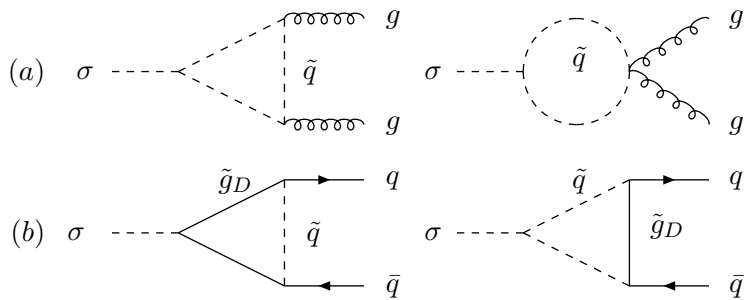
lines, Fig.~\ref{fig:loops}, generate $\sigma gg$ and
$\sigma q \bar q$ couplings.  The
interaction Lagrangian however, eq.~(\ref{eq:sqq}), implies that all $L$- and $R$-squark contributions
to the couplings come with opposite signs so that they cancel each other for
mass degenerate squarks. In addition, the quark-antiquark coupling is
suppressed by the quark mass as evident from general chirality rules.

\section{Sgluon production and decays at the LHC}
Since sgluons have a large color charge they might be more copiously produced than squarks at the LHC. In addition they have quite spectacular decay modes, as discussed below.

\subsection{Sgluon decays}
Sgluons  can  decay into a variety of  different channels that include gluinos, squarks, gluons and quarks.
At tree level the $\sigma$ particles can decay only to a pair of Dirac gluinos
${\tilde{g}}_D$ or into a pair of squarks (if kinematically open),
\begin{eqnarray} \label{eq:tree1}
\Gamma [\sigma \to \tilde{g}_D {\bar{\tilde{g}}}_D ]
   &=& \frac{3 \alpha_s M_{\sigma}}{4}
        \beta_{\tilde{g}}\, (1+\beta^2_{\tilde{g}})\,, \\
\Gamma[ \sigma \to \tilde{q}_a \tilde{q}_a^*] &=& \frac {\alpha_s
   |M_3^D|^2} {4 M_\sigma} \beta_{\tilde{q}_a}\,, \label{eq:tree2}
\end{eqnarray}
where $\beta_{\tilde{g},\tilde{q}_a}$ are the velocities of $\tilde{g},\tilde{q}_a$  ($a=L,R$). If $M_{\sigma} < 2 M_{{\tilde{g}}_D}, 2 m_{\tilde q}$, one or both of these sparticles can be virtual.  Squarks and gluinos  will further cascade to SM particles and LSP.

Decays into gluon or quark-antiquark pairs can proceed only via loop-induced couplings. In the case of no $L/R$ squark mixing\footnote{If $\tilde
q_L$ and $\tilde q_R$ of a given flavor mix, the subscripts $L,R$ have to
be replaced by $1,2$ labeling the squark mass eigenstates, and the
contribution from this flavor is suppressed by the
factor $\cos ( 2 \theta_q)$.} the decay widths take the form

\begin{eqnarray}
\label{loop-gg}
\Gamma(\sigma \rightarrow g g)
  &=& \frac {5 \alpha_s^3} {384 \pi^2} \frac{|M_3^D|^2} {M_\sigma}
      | \sum_q [ \tau_{\tilde{q}_L} f(\tau_{\tilde{q}_L})
                         - \tau_{\tilde{q}_R} f(\tau_{\tilde{q}_R}) ]|^2\,,\\
\Gamma(\sigma \rightarrow q \bar q)
   &=& \frac{9 \alpha_s^3} {128 \pi^2} \frac{|M_3^D|^2 m_q^2}{M_\sigma}\,
     \beta_q
     [ (M^2_\sigma-4 m_q^2) |{\cal I}_S|^2
           +M^2_\sigma \, |{\cal I}_P|^2 ]\,, \label{loop-qq}
\end{eqnarray}
where  $\tau_{\tilde{q}_{L,R}} = 4 m^2_{\tilde q_{L,R}} / M^2_\sigma$ and
$f(\tau)$ is the standard  function from a squark circulating in the  loop~\cite{MS}, while
 the effective scalar (${\cal I}_S$) and pseudoscalar (${\cal I}_P$) $\sigma q\bar q$
couplings  are given by ($i=S,P$)
\begin{eqnarray} \label{I_loop}
{\cal I}_i = \int_0^1 dx \int_0^{1-x} dy [w_i (  C^{-1}_L - C^{-1}_R )
      +z_i (D^{-1}_L - D^{-1}_R) ]
       \,.
\end{eqnarray}
In the above expression $w_S=1-x-y$, $w_P=1$, $z_S=(x+y)/9$, $z_P=0$,  and the squark/gluino denominators
are ($a= L,R$) \\[3mm]
$C_a = (x+y) |M_3^D|^2 + (1-x-y) m^2_{\tilde q_a} - x y M^2_\sigma
 - (x+y) (1-x-y) m_q^2$,\\[1mm] $
D_a = (1-x-y) |M_3^D|^2 + (x+y) m^2_{\tilde q_a}
- x y M^2_\sigma - (x+y) (1-x-y) m_q^2$.\\[-1mm]

Because of the chirality, decays to quark-antiquark pairs are suppressed by the quark mass. Therefore the decays into top quark pairs will dominate.
In addition   both loop-induced decays $\sigma \rightarrow g g$ and $\sigma \rightarrow q \bar q$ are absent if $L$ and $R$ squarks are degenerate. Therefore squarks with substantial mixing (mostly top squarks) will contribute to the decay widths.

The hierarchy between the tree-level and loop-induced
decay modes depends on the values of various soft
breaking parameters.
The tree--level, two--body decays of eqs.~(\ref{eq:tree1},\ref{eq:tree2}) will dominate if they are kinematically allowed. Well above all thresholds the partial width into gluinos always
dominates: it grows $\propto M_\sigma$ since the supersymmetric $\sigma \tilde g \bar{\tilde g}$
coupling is dimensionless, while the partial width into
squarks asymptotically scales like $1/M_\sigma$ due to dimensionfull SUSY-breaking $\sigma \tilde q \tilde q^*$ coupling.

When the above decay channels are shut kinematically, even for small $L$-$R$ squark mass splitting the loop-induced sgluon decays into two gluons, and to a $t \bar t$ pair if
kinematically allowed, always dominate over tree-level off-shell
four--body decays $\sigma \rightarrow \tilde g q \bar q \tilde \chi$  and  $\sigma \rightarrow q \bar q \tilde \chi \tilde \chi$. Increasing the gluino mass increases the $\sigma \tilde q
\tilde q^*$ coupling.  As a result, the partial width into two gluons (due
to pure squark loops) increases, while the $t \bar t$ partial width
(due to mixed squark--gluino loops)  decreases since the increase of the $\sigma \tilde q \tilde q$ couplings is
over--compensated by the gluino mass dependence of the propagators.

\begin{figure}[t]
\vskip 1mm
\begin{center}
\rotatebox{270}{\epsfig{figure=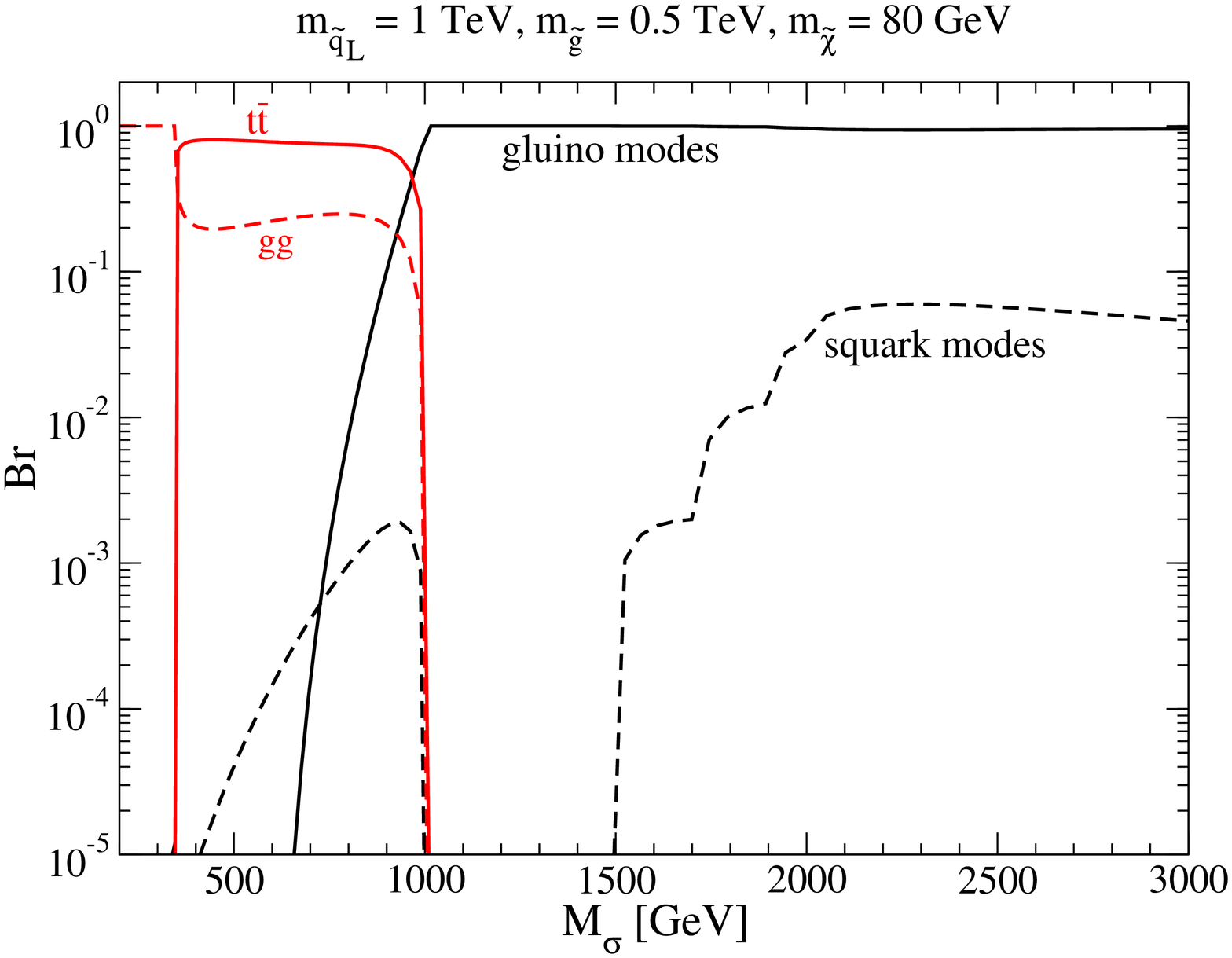, width=5.4cm,height=6cm}}
\rotatebox{270}{\epsfig{figure=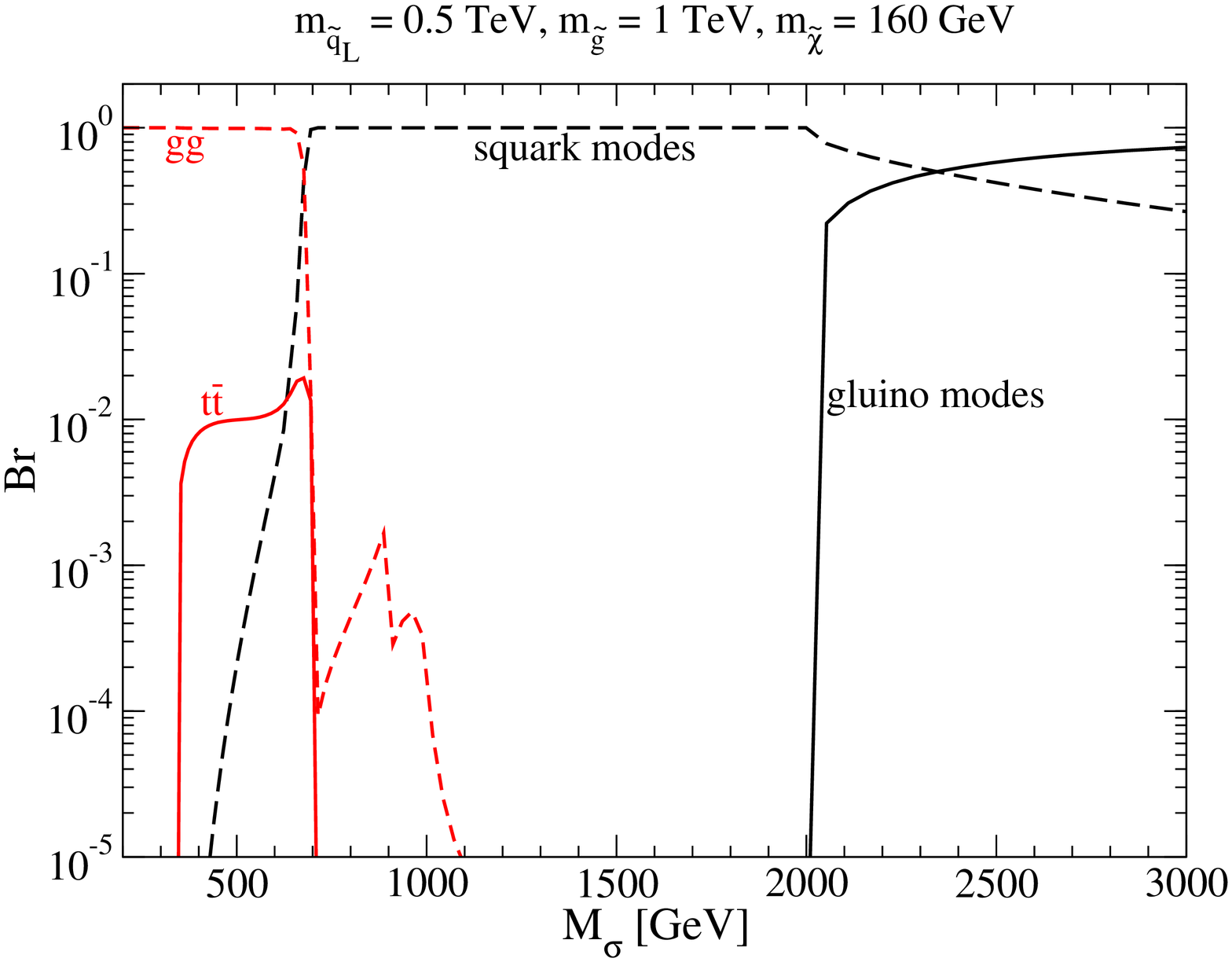, width=5.4cm,height=6cm}}
\caption{\it Branching ratios for $\sigma$ decays, for $m_{\tilde q_L}$=2$
  m_{\tilde g}$=1~TeV (left) and $m_{\tilde g}$ = 2$m_{\tilde q_L}$=1~TeV
  (right). }
\label{fig:br}\end{center}
\end{figure}

The above qualitative features are verified numerically  in Fig.~\ref{fig:br}, where the branching
ratios for $\sigma$ decays are shown for two different squark/gluino mass hierarchies. Plots are done for moderate mass
splitting between the $L$ and $R$ squarks of the five light flavors, and
somewhat greater for soft breaking $\tilde t$ masses: $m_{\tilde q_R} = 0.95
m_{\tilde q_L}, \, m_{\tilde t_L} = 0.9 m_{\tilde q_L},\, m_{\tilde t_R} = 0.8
m_{\tilde q_L}$ with the off--diagonal element of the squared
$\tilde t$ mass matrix $X_t = m_{\tilde q_L}$; the gluino is taken to be a pure Dirac state, i.e. $m_{\tilde g} = |M_3^D|$.

\subsection{Sgluon production at the LHC}
\begin{figure}[t]
\begin{center}
\epsfig{figure=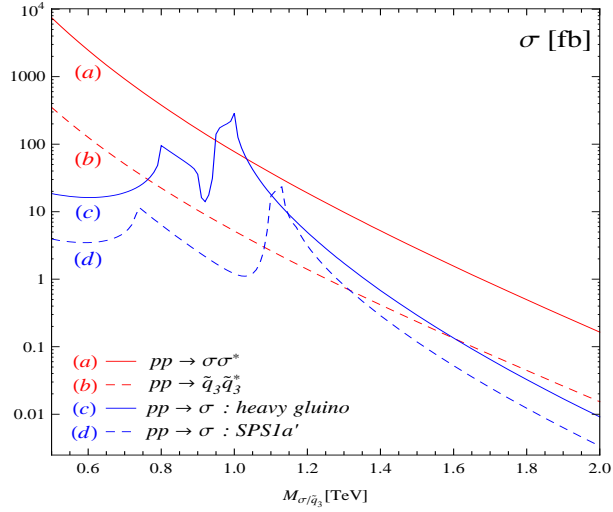, height=6.8cm, width=8cm}
\caption{\it Cross sections for $\sigma$-pair [and $\tilde{q}_3$-pair]
         production (lines (a) and (b)), as well as for single $\sigma$ production (lines (c) and (d)),
         at the LHC. The curve (c) is for the mass parameters as in Fig.3 (right), while the dashed (d) for the mSUGRA benchmark point SPS1a$'$.}
\label{fig:lhcsigma}
\end{center}
\end{figure}

The signatures for single $\sigma$ production are potentially very exciting. They can be produced singly in gluon-gluon collisions
via squark loops since  the production via $q\bar q$ annihilation is negligible for light incoming quarks. The partonic cross section is given by
\begin{equation} \label{sig_gg}
{\hat{\sigma}} [ gg \to \sigma ] = \frac{\pi^2} {M^3_\sigma} \Gamma(\sigma \to
gg)\,,
\end{equation}
where the partial width for $\sigma \to gg$ decays is given in
eq.$\,$(\ref{loop-gg}).
The expected  cross section for single $\sigma$ production  can be quite large at the designed LHC energy, of the order 100 fb. Its dependence on the sgluon mass is shown
by the curves (c) and (d) in Fig.$\,$\ref{fig:lhcsigma} (with the LO CTEQ6L parton
densities \cite{CTEQ}).
In the scenario of the right frame of  Fig.~\ref{fig:br}, the single $\sigma$ cross section  can exceed the $\sigma$-pair
production cross section for $M_\sigma \sim 1$ TeV (solid curve (c)); the dashed one (d) is for  the benchmark point SPS1a$'$ \cite{sps} with the gluino mass interpreted as a Dirac mass.  Since the latter scenario has a
somewhat smaller gluino mass,
it generally leads to smaller cross sections for single $\sigma$ production.

It would be very exciting to observe the sgluon as an $s$-channel resonance.
This however will not be easy since (a) the 2-gluon decay channel must be discriminated from large background,  (b) large production cross section in the gluon fusion implies diminished decay rates  to other channels, which  in addition do not allow   a  direct reconstruction of $M_\sigma$. Detailed experimental simulations are needed to see if the single $\sigma$ production can be
detected as a resonance above the SM and SUSY backgrounds.

Sgluons can also be pair-produced in $q\bar q$ and $gg$ processes at tree-level,
\begin{eqnarray}
\sigma [q\bar{q} \to \sigma\sigma^{\ast} ] &=& \frac{4 \pi \alpha_s^2}{9s}
     \,\beta^3_\sigma \,, \\
\sigma [gg \to \sigma\sigma^{\ast} ]
     &=& \frac{15 \pi \alpha_s^2\beta_\sigma}{8s}
     [ 1 + \frac{34}{5}\, \frac{M_\sigma^2}{s}-\frac{24}{5}
     (1-\frac{M_\sigma^2}{s})\frac{M_\sigma^2}{s}\,
     L_\sigma ]\,,
\end{eqnarray}
where $\sqrt{s}$ is the
invariant parton-parton energy, $M_{\sigma}$ and $\beta_\sigma$ are the mass and center-of-mass velocity of the
$\sigma$ particle, and $L_\sigma =\beta^{-1}_\sigma
     \log(1+\beta_\sigma)/(1-\beta_\sigma)$.
The cross section for $\sigma$-pair production at LHC, $pp \to \sigma
\sigma^\ast$, is shown by the solid curve (a) in Fig.$\,$\ref{fig:lhcsigma}
for the $\sigma$-mass range between  500 GeV and 2 TeV. With
values from several picobarn downwards, a sizable $\sigma\sigma^\ast$ event
rate can be generated. As expected, due to large color charge of the sgluon, the $\sigma\sigma^\ast$ cross section exceeds stop or sbottom-pair
production (dashed line (b)), mediated by a set of topologically equivalent Feynman diagrams,
by more than an order of magnitude.

With the exception of $\sigma\to gg$, all the $\sigma$ decays  give rise to signatures that should easily be detectable. Most spectacular ones come  from $\sigma
\rightarrow \tilde g \tilde g$ decay followed by the gluino decays giving at the end at least
four hard jets and two invisible neutralinos as LSP's. The $\sigma$-pair
production then generates final states with a minimum of four LSP's and eight jets with transverse momenta distributions
markedly different from the corresponding standard  gluino or squark production with the same mass
configurations~\cite{our}.
Other interesting final states  are
four top squarks  $\tilde t_1 \tilde t_1 \tilde t_1^* \tilde t_1^*$,
which can dominate if $m_{\tilde q} \lsim m_{\tilde g}$ and $L$-$R$ mixing is significant
in the stop sector, and $\tilde q \tilde q^* \tilde g \tilde g$ if $M_{\sigma} > 2 m_{\tilde g} \gsim 2 m_{\tilde q}$.
They also give rise to four LSP's in the final state and a large number
of hard jets.

The $gg+gg$ and $t\bar t+ t \bar t$ final states, which
can be the dominant modes if the two--body decays into squarks and gluinos are
kinematically shut, might also allow the direct kinematic reconstruction of
$M_\sigma$. Preliminary analyses of the $gg+gg$ channel show that pairing the four hardest jets in the central region of rapidity  with no missing $E_T$ could reveal the signal and provide a means to reconstruct the $\sigma$ mass directly~\cite{renaud}. In addition, the observation of $t c \bar t \bar c$ final states would signal a substantial mixing in the up-type squark sector.

\section{Summary}
Models with Dirac gauginos offer an interesting alternative to the standard SUSY scenario.
They
predict the existence of scalar particles in the adjoint representation of the gauge groups.
The color-octet  scalars, sgluons,   can be copiously produced at the LHC since they carry large color charge.
Their signatures at the LHC are distinctly
different from the usual  topologies. Depending on the assumed scenario of masses and SUSY breaking parameters, either multi-jet final states with high sphericity and
large missing transverse momentum are predicted, or four top quarks and/or four gluon jets should be
observed in $2\sigma$ production.  In the latter the sgluon mass could be reconstructed directly. If the mass splitting between $L$ and $R$
squarks is not too small, loop--induced single $\sigma$ production may also
have a sizable cross section. It remains to be checked by detailed experimental simulations whether such a single resonant $\sigma$ production is detectable on the large SM and SUSY background.

\section*{Acknowledgments}
We thank Peter Zerwas and Manuel Drees for a fruitful collaboration.
Work supported in parts by the Korea Research Foundation Grant KRF-2008-521-C00069, the Polish Ministry of Science and Higher Education Grant
  N N202 230337, and  the EC Programmes MRTN-CT-
2006-035505 ``Tools and Precision Calculations for Physics Discoveries at Collider'' and MTKD--CT--2005--029466 ``Particle
Physics and Cosmology: the Interface''. JK is grateful  to
the Theory Division for the hospitality extended to him at CERN.


\end{document}